\documentclass[12pt]{article}
\usepackage{graphicx}
\begin{document}
\centerline{\bf Tobin tax and market depth\footnote{We thank Thomas Lux for bringing this group together and for his suggestion to take into account the feedback that reduced speculation via reduced market depth may increase exchange rate variability, and B. Rosenow for comments on the manuscript.}}
\bigskip
\centerline{G. Ehrenstein$^a$, F. Westerhoff$^b$ and D. Stauffer$^a$}

\centerline{$^a$ Institute for Theoretical Physics, Cologne University}

\centerline{Z\"ulpicherstra\ss{}e 77, D-50937 K\"oln, Euroland}

\bigskip
\centerline{$^b$ Department of Economics, University of Osnabr\"uck}

\centerline{Rolandstra\ss{}e 8, D-49069 Osnabr\"uck, Euroland}

\bigskip
\centerline{e-mails: ge@thp.uni-koeln.de, fwesterho@oec.uni-osnabrueck.de, stauffer@thp.uni-koeln.de}

\vspace*{1.5cm}
\begin{abstract}
This paper investigates - on the basis of the Cont-Bouchaud model - whether a Tobin tax can stabilize foreign exchange markets. Compared to earlier studies, this paper explicitly recognizes that a transaction tax-induced reduction in market depth may increase the price responsiveness of a given order. We find that the imposition of a transaction tax may still achieve a triple dividend: (1) exchange rate fluctuations decrease, (2) currencies are less mispriced, and (3) central authorities raise substantial tax revenues. However, if the price impact function is too sensitive with respect to market depth, stabilization may turn into destabilization.

\end{abstract}
\vspace*{1.5cm}

\begin{center}
\textbf{Keywords}
\end{center}
\begin{center}
econophysics; Cont-Bouchaud model; foreign exchange markets;\\ 
Tobin tax; non-linear price impact function; market depth; market efficiency
\end{center}
\newpage


\section{Introduction}
Since the mid 1980s, the daily turnover in financial markets has increased sharply. Moreover, the trading volume increasingly reflects very short-term and speculative transactions. In foreign exchange markets, for example, operations of intraday traders account for 75 percent of the market volume (Bank for International Settlements 2002). In comparison, only 15 percent of the trading volume is on account of non-financial customers, with international trade transactions representing merely 1 percent of the total. The fast and hectic trading leads to complex financial market dynamics. According to Cont (2001) and Lux and Ausloos (2002), the behavior of financial prices may be characterized by five universal features: (1) the evolution of the prices shows little pair correlations between successive daily changes, (2) severe bubbles and crashes occasionally emerge, (3) the prices fluctuate strongly, (4) the distribution of log price changes possesses fat tails, and (5) periods of low volatility alternate with periods of high volatility.\\

Two competing views exist about the efficiency of financial markets. The efficient market hypothesis states that prices reflect their fundamental values. Thus, the statistical features of asset price changes are fully explained by those of the underlying fundamental process. For instance, volatility clustering arises since the intensity of news alternates over time. Extreme price changes reflect the arrival of very important new information. However, it is hard to imagine that the aforementioned stylized facts are fully caused by an exogenous news process.\\

Models with heterogeneous interacting agents seem to describe the working of financial markets more realistically than the traditional neo-classical paradigm. For instance, in Palmer et al. (1994), Kirman (1991), Brock and Hommes (1998), Cont and Bouchaud (2000), Lux and Marchesi (2000), or Farmer and Joshi (2002), the dynamics is mainly driven endogenously through the activity of boundedly rational speculators. Complicated dynamics may arise due to non-linear trading strategies, switching between different types of predictors, or social interactions such as herding behavior. Clearly, these models indicate that financial markets may not be efficient.\\

If the activity of speculators creates distortions, it is interesting to ask whether there exist any means to regulate these markets. Recently, several models with heterogeneous interacting agents have been applied as computer laboratories to explore whether certain policy measures may stabilize financial markets. Note that such simulation experiments have the advantage that they allow the exploration of a certain policy in a well-defined and controlled environment. For instance, one can control for all kinds of random shocks, measure the policy objectives precisely and produce as many observations as required. \\

The focus of this paper is how the Tobin tax affects foreign exchange markets. As early as 1972, Tobin (1978) suggested imposing a uniform tax of around 1 percent on all currency transactions in order to curb speculation. Nowadays, a tax rate of between 0.05 and 0.5 percent is being discussed (Eichengreen et al. 1995, Haq et al. 1996, Frankel 1996, Mende and Menkhoff 2003). Supporters of Tobin's proposal claim that a transaction tax favors long-term investments over short-term investments. Note that around 80 percent of the daily speculation trade takes place because traders would like to take advantage of profits below the $10^{-3}$ border. The effect of a small tax rate could therefore be quite strong. On the other hand, a low tax rate should not harm firms engaged in international trade. Advocates of the Tobin tax also argue that such a device could also raise a substantial amount of tax revenues.\\

Ehrenstein (2003), using the microscopic herding model of Cont and Bouchaud (2000), finds that a Tobin tax may successfully reduce exchange rate volatility. Moreover, the tax revenue in some of the model versions is maximized at a tax rate of around 0.5 percent, which sounds quite realistic. Westerhoff (2003a) develops a simple model with interacting chartists and fundamentalists. He also reports that a small transaction tax may stabilize foreign exchange markets. But if the tax rate is too high, i.e. above 1 percent, too many stabilizing fundamental traders may leave the market and mispricing may increase again. However, both papers have overlooked an important feedback mechanism which may counter the influence of a Tobin tax. The reduction in short-term transactions naturally reduces market depth which may, in turn, increase volatility. Clearly, the price adjustment due to a given order depends on market depth: The less liquid a market is, the stronger the price responsiveness of a given transaction.\\

The aim of this paper is to re-examine the effectiveness of the Tobin tax. We use a modified version of the Cont-Bouchaud model (2000) in which the communication structure between the traders is modeled as a random graph. Cont and Bouchaud show that interactions between market participants through imitation can lead to large fluctuations in aggregate demand. Since the Cont-Bouchaud model gives a reasonable description of financial markets and is able to generate realistic price dynamics we feel safe to use it as a computer laboratory. We find that the imposition of a transaction tax may decrease both volatility and distortions even if a reduction in market depth increases the price responsiveness of a given trade. Moreover, policy makers may raise substantial tax revenues. However, if the price impact function is too sensitive with respect to liquidity, stabilization may turn into destabilization.\\

The paper is organized as follows. Section 2 briefly repeats the basic elements of the Cont-Bouchaud framework. Section 3 presents the experimental design and section 4 summarizes our main results. The final section concludes the paper.    

\section{The model of Cont and Bouchaud}
The goal of Cont and Bouchaud (2000) is to study the impact of herding behavior among speculators on asset price dynamics. Let us briefly repeat the model's main components. We put our agents onto a randomly occupied square lattice\footnote{This model is based on the percolation theory. In the percolation theory we start to fill the lattice such that each site is randomly occupied with probability $p$ and empty with probability $(1-p)$. Neighboring occupied sites form clusters. If a contiguous path of occupied sites connects the top and bottom of the lattice for the first time, the threshold value $p$= $p_c$ is reached.} since previous work (Stauffer 2001) showed that the type of the lattice does not matter much. Cont and Bouchaud consider a stock market with $N$ agents, labeled with an integer $1 \le i \le N$, trading a single asset. During each time period, the agents have three options: to buy one unit of the asset, to sell one unit of the asset, or to remain inactive. The demand of agent $i$ in period $t$ is represented by
\begin{equation}
\label{1}
D_{i}(t)=  \left\{ \begin{array}{r@{\quad:\quad}l}
                        +1 & \mbox{with prob $a$}\\  -1 & \mbox{with prob $a$}\\ 0 & \mbox{otherwise}
                        \end{array}  \right. 
\end{equation}
where the parameter $0 \le a \le 0.5$ captures the activity of the agents. A value of $a < 0.5$ obviously allows for a finite fraction of agents not to trade during a given period\footnote{We can also interpret activity $a$ as a measure of the length of the time we handle in one iteration. If $a$ is close to 0.5, we simulate low frequency data since nearly all market participants are active. Otherwise, a small $a$ reflects high frequency trading since only few agents are active in a given time step.}. In order to focus on the effect of herding, Cont and Bouchaud do not explicitly model the decision process leading to the individual demands. Their random character may, for instance, be due to random resources of the agents. Such behavior is often called noise trading.\\

Aggregate excess demand, i.e. the sum of all orders, is the sole driving force of the asset price: Excess buying drives up the price and excess selling drives down the price. The price adjustment is formalized by a log-linear price impact function
\begin{equation}
\label{2}
P(t+1)= P(t)+ \frac{1}{b}\sum_{i=1}^N{D_{i}(t)},
\end{equation}
where $P(t)$ denotes the log price at time $t$ and $b$ stands for a positive liquidity parameter describing how much excess demand is needed to move the asset price by one unit. Note that log price changes and excess demand vary proportionally. Cont and Bouchaud set $b$= 1.\\

In real markets, agents may form groups of various sizes which may then share information and act in coordination\footnote{If the orders of the agents were independent then the returns would be normally distributed. However, this is not consistent with the data.}. The agents' group formation is described through a random matching process. All agents which are direct or indirect neighbours of each
other form a cluster which adopts one common strategy of selling or buying. 
Each agent has at most four direct neighbours but a large cluster (= company
or coalition) can be formed through the neighbours of neighbours etc. For 
an occupation probability $p$ above $p_c = 0.592746$ a cluster connecting
top and bottom is formed; we work at this critical concentration.\\

The microscopic model of Cont and Bouchaud and its variants have the power to mimic actual asset price dynamics quite closely (Stauffer 2001). For an activity $a$ close to 0.5 and $p$= $p_c$ the distribution of the returns is similar to a Gaussian curve. However, for a smaller activity level, one obtains heavy tails in the distribution of the returns. Moreover, weak correlations exist between successive returns and strong correlations between successive absolute returns. Since prices do not react to news, one may also argue that the speculators cause distortions and excess volatility.\\

\section{Laboratory design}
\subsection{Modifications}
Some adjustments are necessary to be able to study the effectiveness of the Tobin tax within the Cont-Bouchaud framework.\\

First, since log-price changes of the Cont-Bouchaud model are often large integers and the Tobin tax is a very small number (i.e. less than 1 percent), we have to normalize the returns. In reality, extreme price changes in major foreign exchange markets seldomly exceed the 5 percent level. Thus, we take here $maxwin=5$ percent. This means that if all clusters in an iteration are active and buying, the return is +5 percent. Otherwise, if all clusters are active and selling, the return is set to -5 percent. Certainly, not all clusters will trade in the same direction in the same iteration.\\

Second, we impose a Tobin tax on all currency transactions. This changes the behavior of the speculators in the following way. The speculators believe that the log-price change in period $t-1$ is authoritative for the log-price change development in period $t$. Thus, if the absolute value of the log-price change is lower than the tax rate, speculation is identified as not profitable. As a result, speculators become inactive.\\

Third, we include international trade transactions. The orders of international firms, on which the Tobin tax has no impact, consist of two elements: An unsystematic random component and a systematic deterministic component. The unsystematic component reflects random liquidity needs of the firms, e.g. to pay bills in a foreign currency. This is implemented by assuming that 1 percent of all clusters describe the behavior of firms. The systematic component is due to current account imbalances. For instance, if the exchange rate is overvalued then exports exceed imports. The systematic demand of the international firms is given as
\begin{equation}
\label{3}
\Delta(t)=(F-P(t-1))d,
\end{equation}
where $F$ is the log of the fundamental value and $d$ is a positive reaction coefficient. According to (3), current account imbalances increase with the mispricing of the exchange rate. We assume that $F=0$ and $d=0.001$, which implies that each day $1/10$ of 1 percent of any gap from the fundamental value dissipates. This translates into a realistic half-life of about 2 years. \\

Fourth, Cont and Bouchaud assume a proportionality between aggregated excess demand and log-price changes which is a reasonable approximation as long as the market depth does not vary too strongly. However, since the Tobin tax may significantly crowd out speculative transactions, we introduce a non-linear price impact function. In particular, we assume that a given transaction causes a small (large) price change if market liquidity is high (low). Let $e$ be the normalization factor to scale the dynamics according to $maxwin$, then the price adjustment may be written as
\begin{equation}
\label{4}
 P(t+1)=P(t)+A(\tau,t)(\Delta(t) + e \sum_{i=1}^N D_i(t)) 
\end{equation}
with
\begin{equation}
\label{5}
A(\tau,t)=\frac{f}{[\sum_{k=1}^\tau{(\mid \Delta(t-k+1)\mid+ \sum_{i=1}^N{\mid D_{i}(t-k+1)\mid)]^g}
}}.
\end{equation} 
\\
The exponent $g \ge 0$ captures the curvature of the price adjustment while $f$ is a positive shift parameter. The market depth is given as the sum of all currency transactions within the last $\tau$ trading periods. Note that for $g=0$ and $f=1/b$, (4) is identical to the price impact function of the original Cont-Bouchaud model. For $g > 0$, the price impact of a given order decreases with increasing liquidity\footnote{The non-linearity of the price impact function may create chaotic price dynamics even if the behavior of speculators is deterministic and linear (Westerhoff 2003b).}.

\subsection{The algorithm}
In a nutshell, the simulations are executed as follows:

\begin{itemize}
\item With the algorithm of Hoshen and Kopelman we determine 
the number of clusters with $s$ agents.
\item We decide randomly if the cluster is active in this iteration.
\begin{itemize}
\item If the cluster is active we test whether the condition for profitable speculation is fulfilled.
\begin{itemize}
\item If this is the case we decide by another random number if the cluster would like to buy or sell an amount which corresponds to the size of the cluster.
\item If the condition is not true we decide through another random number if the cluster is forced to trade because it belongs to one of the international firms.
\begin{itemize}
\item If the cluster is an international firm, we decide randomly whether the cluster would like to buy or sell.
\item If the cluster does not trade it has no impact on the dynamics.
\end{itemize}
\end{itemize}
\item If the cluster is not active it has no impact on the dynamics.
\end{itemize}
\item If all clusters have been processed we determine the new price and the iteration is finished.
\item The procedure is repeated for the next iteration.
\end{itemize}

\subsection {Policy objectives}
Before we turn to the simulation results let us first define three important policy objectives. A high exchange rate variability implies a high risk for internationally operating firms. As is well known, the higher the exchange rate volatility is, the more strongly risk-averse firms retreat from international trade, which is bad for the markets. Thus, policy makers have an incentive to control exchange rate risk. If the exchange rate is misaligned, long-term capital investments may flow into inefficient sectors. To achieve a good capital allocation, prices should reflect their fundamental values closely. Finally, the imposition of a transaction tax generates an additional source of income. We formalize these criteria as follows. Volatility is computed as
\begin{equation}
\label{6}
\mbox{volatility}=\frac{1}{T}\sum_{t=1}^T{\mid P(t)-P(t-1)\mid},
\end{equation}
distortion as
\begin{equation}
\label{7}
\mbox{distortion}=\frac{1}{T} \sum_{t=1}^T{\mid P(t)-F\mid},
\end{equation}
and the tax revenue as
\begin{equation}
\label{8}
\mbox{revenue}= \mbox{tax} [\sum_{t=1}^T(\mid \Delta(t)\mid + \sum_{i=1}^N \mid D_{i}(t)\mid)],
\end{equation}
where $T$ is the sample length and tax is the tax rate. The simulations are based on T=100,000.

\section{Results}
We are now ready to explore the impact of the Tobin tax on the dynamics of foreign exchange markets. We fix the following parameters: $N$= 570, $a$= 0.4999, $p_c$= 0.592746, $d$= 0.001, $f$= 1. Figure 1 shows the results for $\tau=1$, i.e. market liquidity only depends on the actual trading volume. The first panel of figure 1 displays the reaction of the volatility for $g$= 0 (the "+ + +"-line), $g$= 0.19 (the "$\times$ $\times$ $\times$"-line) and $g$= 0.4 (the "$\star$ $\star$ $\star$"-line). The transaction tax is increased from 0 to 1 percent. As can be seen, volatility decreases, remains constant, or even increases due to currency taxation\footnote{Note that the three curves do not start at the same point, i.e. $g$ has an impact on the absolute level of the volatility. However, the shift parameter $f$ allows the rescaling of the starting point such that the curves have the same origin. Such rescaling does not qualitatively change our results.}. The second panel of figure 1 reveals similar results for the distortion. A Tobin tax may help drive prices closer towards fundamentals as long as $g$ is not too large. For instance, for $g$= 0, a Tobin tax of 0.2 percent decreases volatility by more than 50 percent and distortion by around 33 percent. Finally, the third panel of figure 1 presents the income generating potential of the Tobin tax. Up to around $g$= 0.19, the tax revenue function has a maximum. For higher values of the exponent, tax revenues increase with increasing tax rates, at least for tax rates below 1 percent. Note also that the revenue maximizing tax rate may not coincide with the volatility minimizing tax rate (e.g. for $g$= 0). In this sense, policy makers may set the tax rate too low in order to generate higher tax revenues.\\

These results deserve further attention. Note first that the Tobin tax may indeed achieve a triple dividend even if a reduction in market depth increases the price responsiveness of a given order. This may be regarded as good news for policy makers since it allows them to stabilize foreign exchange markets and to generate government income. Proponents of the Tobin tax always have this case in mind. However, our simulations also give a warning to policy makers: The success of a Tobin tax is not absolutely sure. If the curvature of the price impact function is too extreme (i.e. $g>$ 0.2), the Tobin tax destabilizes the market in the sense that both volatility and distortion increase. For $g$= 0.4, even a minimal tax rate always increases exchange rate variability. The latter result stands in sharp contrast to earlier findings on the usefulness of transaction taxes.\\

Are these estimates robust? Figure 2 displays the results for $\tau$= 20, that is the market depth is taken as the trading volume over the last 20 observations (now the "+ + +"-line stands for $g$= 0, the "$\times$ $\times$ $\times$"-line for $g$= 0.2 and the "$\star$ $\star$ $\star$"-line for $g$= 0.4). Since the model refers to daily data, 20 observations correspond to a time span of one month. Again, we find that the Tobin tax is not always stabilizing. However, if market liquidity depends on a longer time horizon, then the advocates of the Tobin tax have reason to be more optimistic. For $g$= 0.2, for instance, we still see a sharp drop in volatility and distortion. Further examinations revealed that as $\tau$ increases further, say up to 50 or 100 trading periods, volatility and distortion decrease for much higher values of the exponent $g$.\\

\section{Conclusions}
Short-term speculations generate excess volatility. As a result, financial markets often lack anchoring in fundamentals. Tobin (1978) thus proposed a levy on all foreign-exchange transactions. The tax should be small enough to be fairly negligible for firms engaged in international trade, yet wipe out a lot of short-term speculation. Short-term financial round-trip excursions amplify even a very low tax rate. For instance, a tax of 0.1 percent, measured in terms of annualized expected rates of return, would come to a 43 percent\footnote{$1.001^{360}=1.433$} penalty on one-day speculation. Although the Tobin tax is frequently discussed in the popular media, it has remained under-researched in academia.\\

This paper uses the well-known herding model of Cont and Bouchaud to investigate the consequences of a transaction tax on foreign exchange dynamics. In contrast to previous studies, this paper takes into account that a reduction in market depth increases the price responsiveness of a given trade. Overall, we find that a transaction tax may help dampen economically unjustified speculation. To be precise, a triple dividend may be achieved: volatility and distortion decrease while government income increases. However, there exist critical values of the exponent $g$ above which market stability may decrease. \\

It may therefore be theoretically possible that a Tobin tax worsens market efficiency. Whether this outcome is realistic or not is an empirical question. So, how strong is the curvature of the price impact function with respect to market depth? Unfortunately, no clear answer exists. Kempf and Korn (1999), using data on DAX futures, and Plerou et al. (2002), using data on the 116 most-frequently traded US stocks, find that the price impact function displays a concave curvature with increasing order size, and flattening at larger values. Put differently, the smaller is the average price impact per trade unit, the larger is the order size. Weber and Rosenow (2003) also fitted a concave function in the
form of a power law and obtained a correlation coefficient of 0.977. The
implications of such price impact functions have earlier been discussed by
Zhang (1999). More closely related with our setup, Lillo, Farmer and
Mantenga (2003) report that higher capitalization stocks tend to have
smaller price responses for the same (normalized) transaction size. Since
market capitalization is correlated with liquidity, the price impact
function is presumbly non-linear. But the non-linearity may not be very
extreme, at least in the case of foreign exchange markets. For instance,
Evans and Lyons (2002), who estimate that \$ 1 billion of net dollar purchases
increases the deutsche mark price of a dollar by 0.5 percent, could not
improve their fit by including non-linarities. However, their data set only
includes about 80 daily observations. Note also that market depth in foreign
exchange markets has increased sharply since the 1970s without producing lower volatility. Put differently, if the current average daily turnover of \$ 1,200 billion would decrease by, say 50 percent, due to currency taxation, market depth would still remain extremely high. A turnover of \$ 600 billion - still higher than the turnover in 1989 - times a tax rate of, say 0.1 percent, would then generate an annual tax revenue of \$ 220 billion. Our simulations ignore the administrative costs of collecting the tax as well as the dangers arising from more government control over its citizens through the tax information.\\[0.5cm]


\textbf{References}\\
Bank for International Settlements (2002): Central Bank Survey of Foreign Exchange and Derivatives Market Activity 2001, Basle.\\[0.1cm]
Brock, W. and Hommes. C. (1998): Heterogeneous Beliefs and Routes to Chaos in a Simple Asset Pricing Model, Journal of Economic Dynamics and Control, 22, 1235-1274. \\[0.1cm]
Cont, R. (2001): Empirical Properties of Asset Returns: Stylized Facts and Statistical Issues, Quantitative Finance, 1, 223-236.\\[0.1cm]
Cont, R. and Bouchaud, J-P. (2000): Herd Behavior and Aggregate Fluctuations in Financial Markets, Macroeconomic Dynamics, 4, 170-196.\\[0.1cm]
Ehrenstein, G. (2003): Cont-Bouchaud Percolation Model Including Tobin Tax, International Journal of Modern Physics C, 13, 1323-1331.\\[0.1cm]
Eichengreen, B., Tobin, J. and Wyplosz, C. (1995): Two Cases for Sand in the Wheels of International Finance, Economic Journal, 105, 162-172.\\[0.1cm]
Evans, M. and Lyons, R. (2002): Order Flow and Exchange Rate Dynamics, Journal of Political Economy, 110, 170-180.\\[0.1cm]
Farmer, D. and Joshi, S. (2002): The Price Dynamics of Common Trading Strategies, Journal of Economic Behavior and Organization, 49, 149-171.\\[0.1cm]
Frankel, J. (1996): Recent Exchange-Rate Experience and Proposals for Reform, American Economic Review, 86, 15-158. \\[0.1cm]
Haq, M., Kaul, I. and Grunberg, I. (1996): The Tobin Tax: Coping with Financial Volatility, Oxford University Press: New York.\\[0.1cm]
Kempf, A. and Korn, O. (1999): Market Depth and Order Size, Journal of Financial Markets, 2, 29-48.\\[0.1cm]
Kirman, A. (1991): Epidemics of Opinion and Speculative Bubbles in Financial Markets, in: Taylor, M. (Ed.): Money and Financial Markets, Blackwell: Oxford, 354-368.\\[0.1cm]
Lillo, F., Farmer, D. and Mantenga, R. (2003): Master Curve for Price Impact Function, Nature, 421, 129-130.\\[0.1cm]
Lux, T. and Marchesi, M. (2000): Volatility Clustering in Financial Markets: A Micro-Simulation of Interacting Agents, International Journal of Theoretical and Applied Finance, 3, 675-702.\\[0.1cm]
Lux, T. and Ausloos, M. (2002): Market Fluctuations I: Scaling, Multiscaling, and Their Possible Origins. In: Bunde, A., Kropp, J. and Schellnhuber, H. (Eds.): Science of Disaster: Climate Disruptions, Heart Attacks, and Market Crashes. Springer, Berlin, 373-410.\\[0.1cm]
Mende, A. and Menkhoff, L. (2003): Tobin Tax Effects Seen from the Foreign Exchange Market's Microstructure, International Finance, 6, 227-247.\\[0.1cm]
Palmer, R., Arthur, B., Holland, J., LeBaron, B. and Tayler, P. (1994): Artificial Economic Life: A Simple Model of a Stock Market, Physica D, 75, 264-274.\\[0.1cm]
Plerou, V., Gopikrishnan, P., Gabaix, X. and Stanley, H.E. (2002): Quantifying Stock-Price response to Demand Fluctuations, Physical Review E, 66, 027104[1]-027104[4]\\[0.1cm]
Stauffer, D. (2001): Percolation Models for Financial Market Dynamics, Advances in Complex Systems, 4, 19-27.\\[0.1cm]
Tobin, J. (1978): A Proposal for International Monetary Reform, Eastern Economic Journal, 4, 153-159.\\[0.1cm]
Weber, P. and Rosenow, B. (2003): Order Book Approach to Price Impact, Preprint cond-mat/0311457\\[0.1cm]
Westerhoff, F. (2003a): Heterogeneous Traders and the Tobin Tax, Journal of Evolutionary Economics, 13, 53-70\\[0.1cm]
Westerhoff, F. (2003b): Market Depth and Price Dynamics, Discussion Paper University of Osnabrueck, WP 2003/15.\\[0.1cm]
Zhang, Y.C. (1999): Toward a theory of marginally efficient markets, Physica A, 269, 30-44

\newpage
\textbf{Legends for figures}\\
Figure 1: The effect of the Tobin tax for $\tau$= 1. The first, second, and  third panels show the volatility, the distortion and the tax revenue as a function of the Tobin tax for different $g$, respectively. The Tobin tax is increased from 0 to 1 percent. The "+ + +"-line, the "$\times$ $\times$ $\times$"-line and the "$\star$ $\star$ $\star$"-line stand for $g$= 0, $g$= 0.19 and $g$= 0.4, respectively.\\[1cm]
Figure 2: The effect of the Tobin tax for $\tau$= 20. The first, second, and third panels show the volatility, the distortion and the tax revenue as a function of the Tobin tax for different $g$, respectively. The Tobin tax is increased from 0 to 1 percent. The "+ + +"-line, the "$\times$ $\times$ $\times$"-line and the "$\star$ $\star$ $\star$"-line stand for $g$= 0, $g$= 0.2 and $g$= 0.4, respectively.

\newpage

\begin{figure}[hbt]
\begin{center}
\includegraphics[angle=-90,scale=0.35]{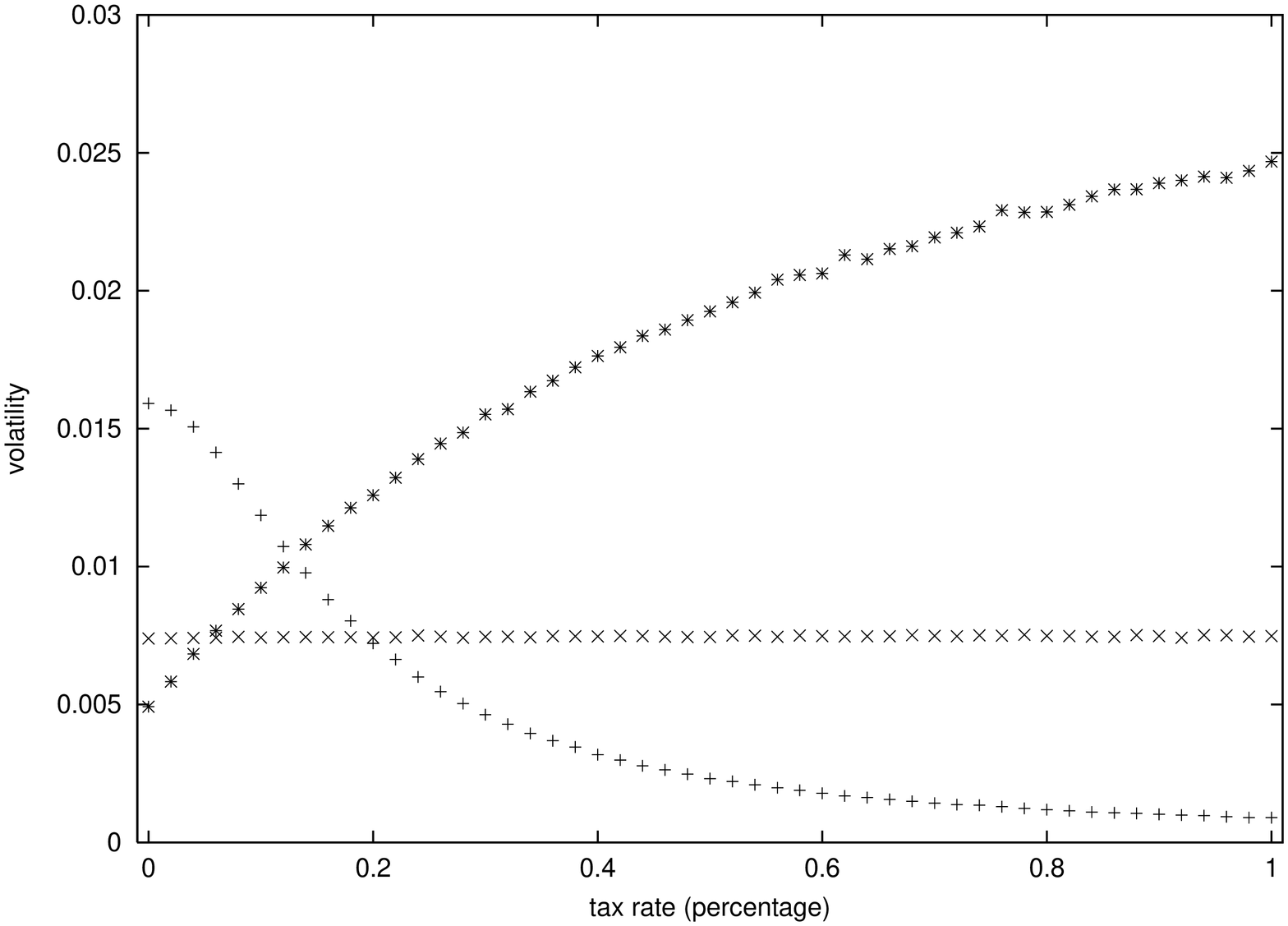}
\includegraphics[angle=-90,scale=0.35]{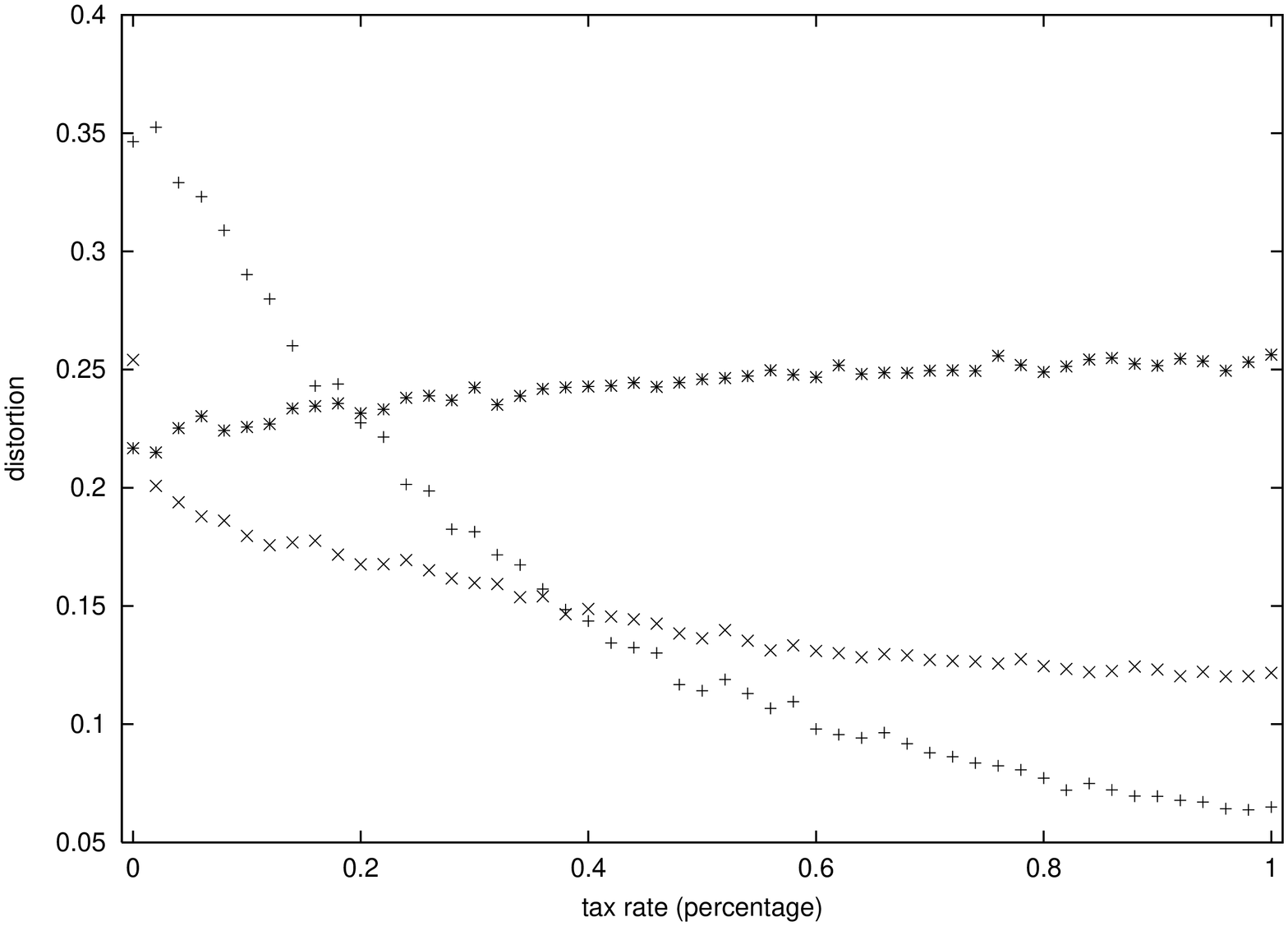}
\includegraphics[angle=-90,scale=0.35]{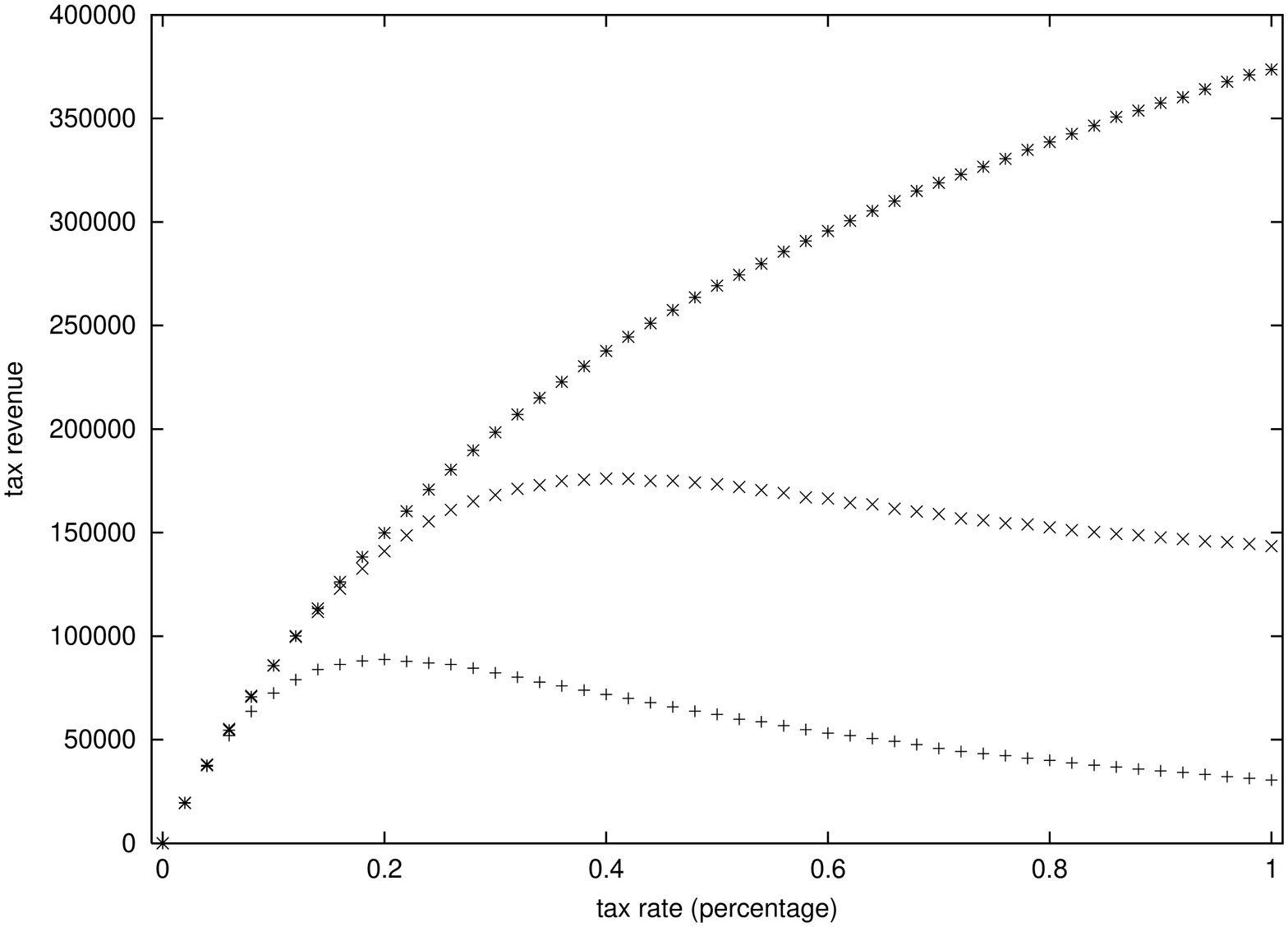}
\end{center} 
\end{figure}

\newpage
\begin{figure}[hbt]
\begin{center}
\includegraphics[angle=-90,scale=0.35]{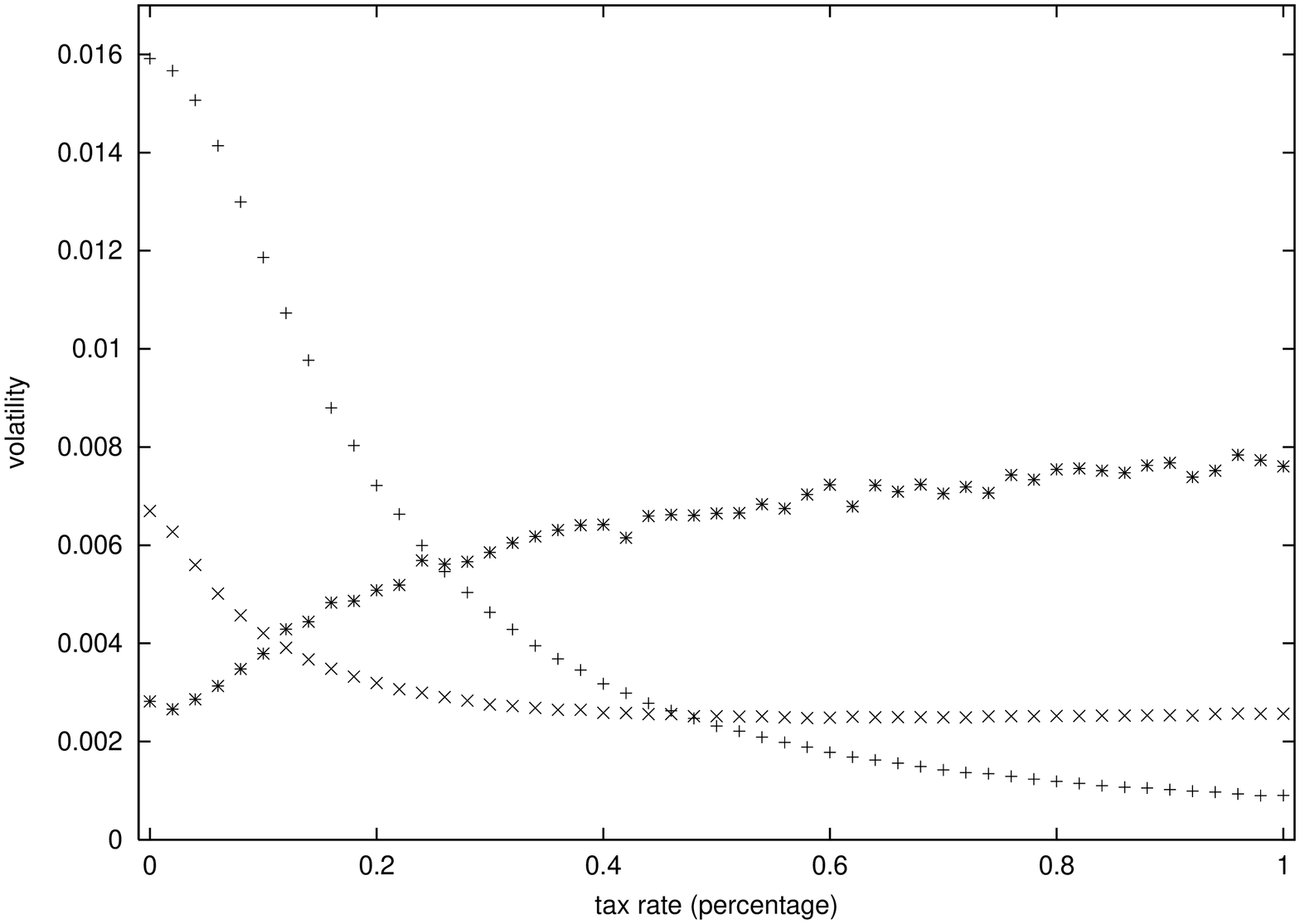}
\includegraphics[angle=-90,scale=0.35]{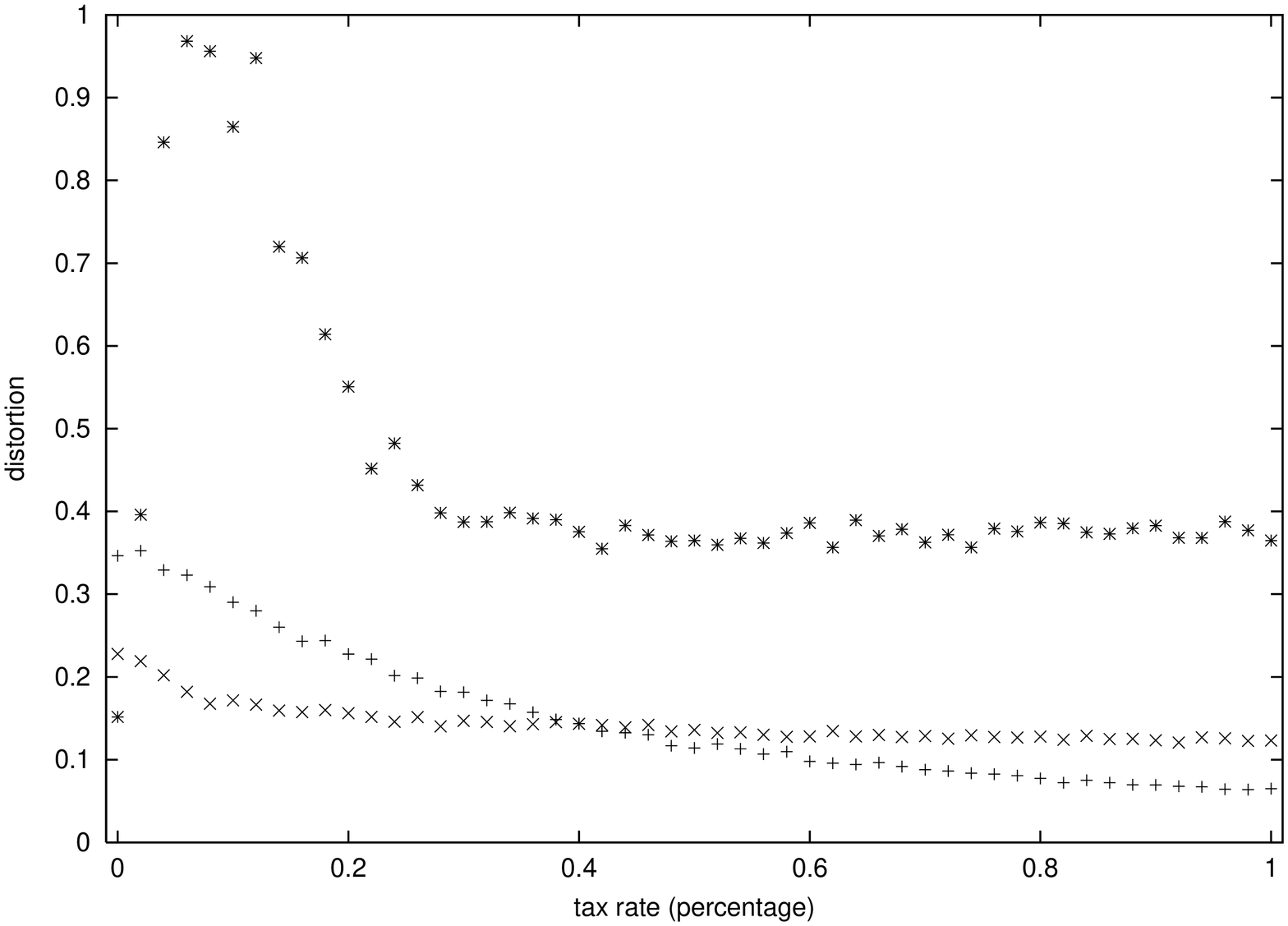}
\includegraphics[angle=-90,scale=0.35]{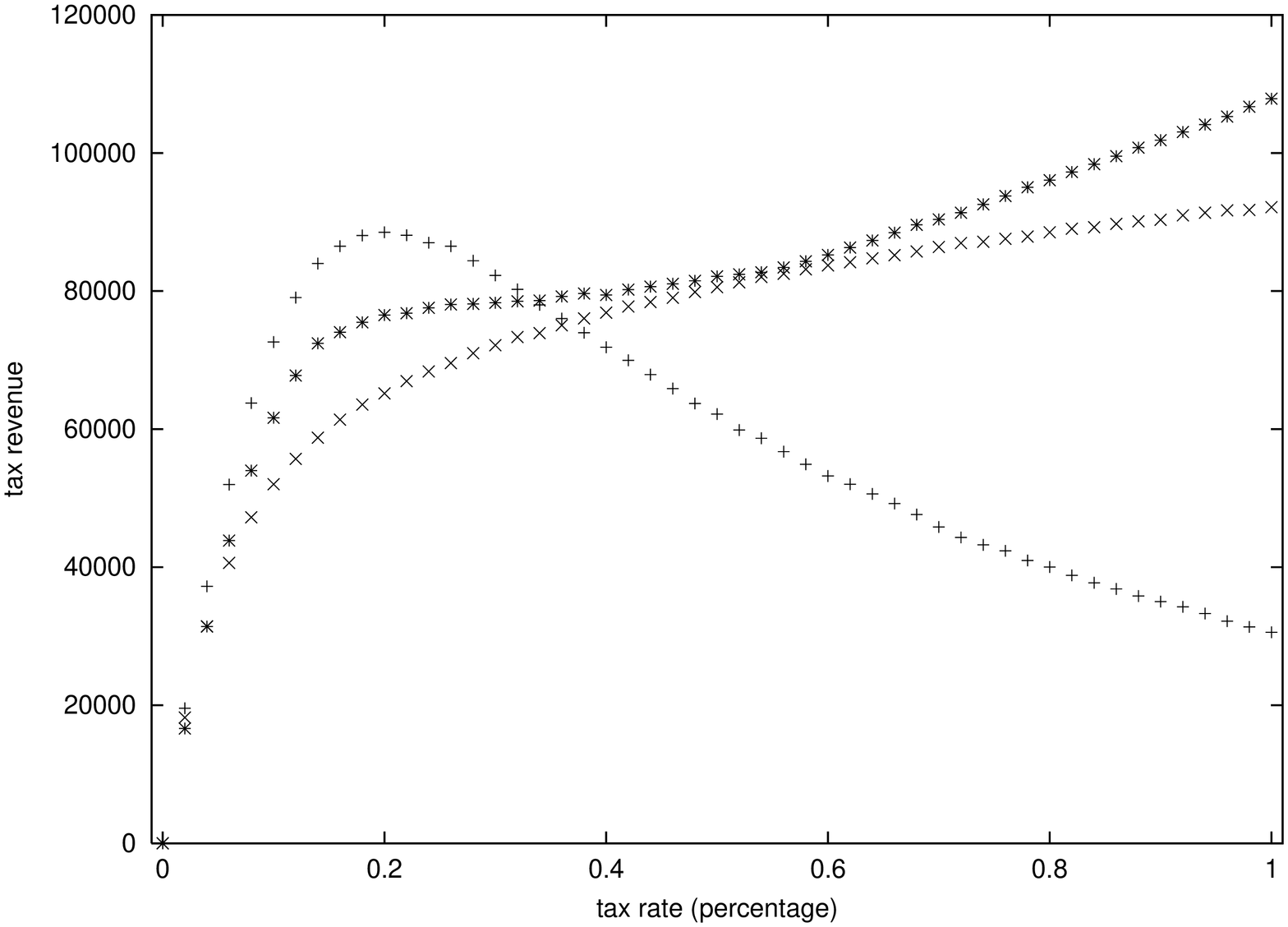}
\end{center} 
\end{figure}

\end{document}